\documentclass[aps,prl,reprint,amsmath,showkeys,nofootinbib]{revtex4-2}
\usepackage[colorlinks,allcolors=blue]{hyperref}
\usepackage[T1]{fontenc}
\usepackage{bm}
\usepackage[nopatch=item]{microtype}
\usepackage{graphicx}
\usepackage{tikzexternal}

\begin{document}

\title{Hidden-Heavy Pentaquarks and Where to Find Them}
\author{Fareed Alasiri}
\email{alasiri.6@osu.edu}
\affiliation{Department of Physics, The Ohio State University, Columbus, Ohio 43210, USA}
\author{Eric Braaten}
\email{braaten.1@osu.edu}
\affiliation{Department of Physics, The Ohio State University, Columbus, Ohio 43210, USA}
\author{Roberto Bruschini}
\email{bruschini.1@osu.edu}
\affiliation{Department of Physics, The Ohio State University, Columbus, Ohio 43210, USA}

\begin{abstract}
We provide a simple explanation for the observed hidden-charm pentaquarks as bound states in Born-Oppenheimer potentials.
We identify $P_{c\bar{c}}(4312)^+$, $P_{c\bar{c}}(4440)^+$, and $P_{c\bar{c}}(4457)^+$ as heavy-quark spin states in a quartet of $c \bar{c}$ pentaquarks with $J^P$ quantum numbers $\frac{1}{2}^-$, $\frac{3}{2}^-$, and $\frac{5}{2}^-$.
The quantum numbers of $P_{c\bar{c}}(4457)^+$ differ from most previous predictions.
We also predict a fourth $c\bar{c}$ pentaquark with quantum numbers $\frac{3}{2}^-$ near the $\Sigma_c^\ast\bar{D}$ threshold.
We identify $P_{c\bar{c}s}(4338)^0$ and $P_{c\bar{c}s}(4459)^0$ as heavy-quark spin states in a triplet of $c \bar{c}s$ pentaquarks with quantum numbers $\frac{1}{2}^-$ and either $\frac{1}{2}^-$ or $\frac{3}{2}^-$.
We also predict a third $c\bar{c}s$ pentaquark with quantum numbers either $\frac{3}{2}^-$ or $\frac{1}{2}^-$ near the $\Xi_c\bar{D}^\ast$ threshold.
We explain why the observed hidden-charm pentaquarks have narrow widths.
\end{abstract}


\maketitle

\textbf{Introduction.}
The unexpected discoveries of dozens of exotic heavy hadrons since 2003 has provided a major challenge to our understanding of quantum chromodynamics (QCD) \cite{Ali17,Esp17,Leb17,Guo18,Kar18,Ols18,Bra20,Husk20,Che23}.
Constituent models, such as molecular models with color-singlet hadron constituents and quark and diquark models with colored constituents, have failed to reveal the pattern of the  exotic heavy hadrons.
The direct assault on this problem using lattice QCD seems to be prohibitively difficult because of their many decay channels.

A promising approach within QCD is  the Born-Oppen\-heimer (B\nobreakdash-O) approximation, which exploits the large mass of a heavy (charm or bottom) quark compared to the energies of gluons and light quarks.
The  B\nobreakdash-O approximation for QCD was introduced by Juge, Kuti, and Morningstar, who applied it to quarkonium hybrids in 1999 \cite{Jug99}.
It separates the problem into two steps: (1) the calculation of B\nobreakdash-O potentials, which are discrete energy levels of QCD in the presence of static color sources as functions of the separation of the sources $r$, and (2) the solution of the Schr\"odinger equation for heavy quarks and antiquarks interacting through those potentials.
This approach has been developed into an effective field theory called BOEFT that can be applied to all multi-heavy hadrons \cite{Bra18a,Sot20a,Ber24}.
However, the development of BOEFT has not revealed the pattern of the exotic heavy hadrons.

A pattern for the exotic hidden-heavy hadrons, which contain a heavy quark and antiquark, was proposed in Ref.~\cite{Braa24b}.
BOEFT implies that their B\nobreakdash-O potentials at short distances are repulsive color-Coulomb potentials offset by the energy of an $\bm{8}$-hadron, which is a discrete energy level of QCD in the presence of a static color-octet source.
Confinement implies that their B\nobreakdash-O potentials at long distances approach thresholds for pairs of heavy hadrons.
Such potentials can support bound states only if they cross below the heavy-hadron-pair threshold before approaching it; see Figure~\ref{potentials}.
The pattern proposed in Ref.~\cite{Braa24b} is that the exotic hidden-heavy hadrons are bound states and resonances in such B\nobreakdash-O potentials.
In Ref.~\cite{Braa24b}, this pattern was applied to nonstrange hidden-heavy tetraquarks and some of their properties were postdicted.
The LHCb collaboration has discovered three nonstrange hidden-charm ($c\bar{c}$) pentaquarks \cite{Aai15,LHCb19,LHCb21} and two strange hidden-charm ($c\bar{c}s$) pentaquarks \cite{LHCb20,LHCb22}.%
\footnote{%
We disregard the broad $c\bar{c}$ pentaquark state $P_{c\bar{c}}(4380)^+$ in the Particle Listings of the Particle Data Group for the reasons indicated in the PDG Review on Pentaquarks \cite{PDG25}.%
}
In this paper, we apply the pattern in Ref.~\cite{Braa24b} to hidden-heavy pentaquarks and we predict some of their properties.

\begin{figure}
\includegraphics[width=\columnwidth]{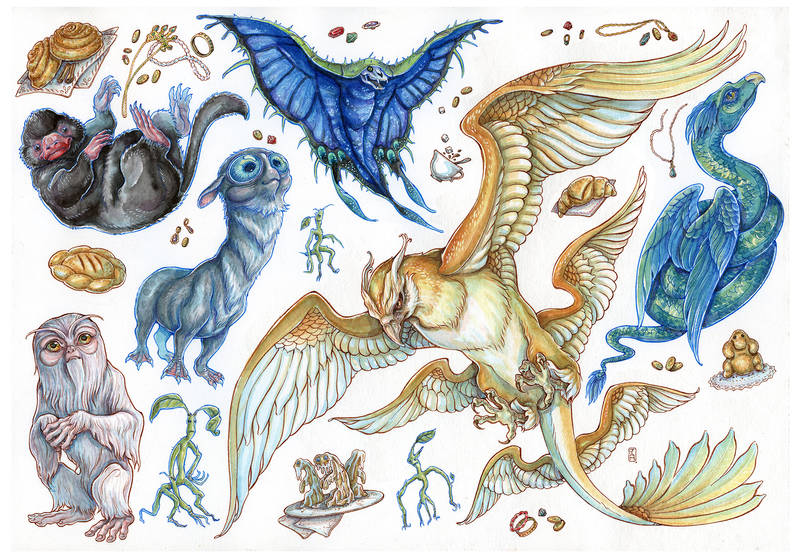}
\caption{Exotic hidden-heavy hadrons and other fantastic beasts (image by Jam-Di, \href{https://www.deviantart.com/jam-di}{deviantart.com/jam-di}).}
\label{fig:beasts}
\end{figure}

\begin{figure}
\centering
\tikzsetnextfilename{potentials}
\begin{tikzpicture}
\pgfmathsetmacro{\hbar}{197.3}
\pgfmathsetmacro{\keight}{0.037}
\pgfmathsetmacro{\sommer}{0.5}
	\begin{axis}
		[width=7.5cm,
		xlabel={$r$},
		ylabel={$V$},
		ylabel style={rotate=-90},
		xmajorticks=false,
		ymajorticks=false,
		axis y line=left,
		axis x line=middle,
		enlarge x limits=.02,
		enlarge y limits=.05,
		colormap/Dark2]
			\addplot
			[index of colormap=0,
			samples=100,
			domain=0.03:3,
			very thick,
			mark=none,
			name path=lower curve]
			{(\keight * \hbar / x - 100) * exp(-x / \sommer)};
			\path[name path=bound state]
			(0,-10) -- (3, -10);
			\draw [index of colormap=0, name intersections={of=lower curve and bound state}]
			(intersection-1) -- (intersection-2);
	\end{axis}
\end{tikzpicture}
\caption{\label{potentials} A B-O potential for exotic hidden-heavy hadrons (thick curve) may support bound states (thin horizontal line) only if it crosses below the heavy-hadron-pair threshold before approaching it.}
\end{figure}

\textbf{Static and Heavy Hadrons.}
A static hadron is a discrete energy level of QCD bound to a static color source.
It is called a $\bm{3}$-hadron, $\bm{\bar{3}}$-hadron, or $\bm{8}$-hadron if the color source is a triplet, antitriplet, or octet, respectively.
Static hadrons are labeled by quantum numbers $j^\pi$ for angular momentum and parity.
In QCD with two light quarks and a heavier strange quark,
static hadrons form degenerate SU(2) multiplets labeled by $(I,S)$ for isospin and strangeness and they form approximately degenerate SU(3) multiplets labeled by their multiplicity.

A $\bm{\bar{3}}$-hadron with a single light quark is a $\bm{\bar{3}}$-meson.
A $\bm{3}$-hadron with two light quarks is a $\bm{3}$-baryon.
At large $r$, a pentaquark B-O potential approaches the threshold given by the sum of the energies of a $\bm{\bar{3}}$-meson and a $\bm{3}$-baryon.
The color, spin, and flavor states of the quarks in a static hadron are in the fundamental representations of SU(3), SU(2), and SU(3), respectively.
The ground-state $\bm{\bar{3}}$-mesons are $S$-wave states with even parity ($\pi=+$).
They therefore form a $j^\pi=\frac{1}{2}^+$ SU(3) triplet consisting of $(I,S)=\bigl(\frac{1}{2},0\bigr)$ and $(0, -1)$.
The state of a $\bm{3}$-baryon must be antisymmetric under exchange of the two quarks.
The ground-state $\bm{3}$-baryons are symmetric in the quark positions and therefore have $\pi=+$.
The color state of the two quarks must be the antisymmetric antitriplet,
so their spin-flavor state must be symmetric.
The ground-state $\bm{3}$-baryons therefore form two SU(3) multiplets: a $j^\pi=0^+$ antitriplet consisting of $(I,S)=(0, 0)$ and  $\bigl(\frac{1}{2},-1\bigr)$ and a $1^+$ sextet consisting of $(1,0)$, $\bigl(\frac{1}{2},-1\bigr)$, and $(0,-2)$.

A heavy hadron contains a single heavy quark or antiquark and light QCD fields that can be approximated by a $\bm{3}$-hadron or $\bm{\bar{3}}$-hadron.
The heavy-quark spin is conserved up to terms that scale like $1/m_Q$ with $m_Q$ the heavy-quark mass.
Heavy hadrons form approximately degenerate heavy-quark spin-symmetry (HQSS) multiplets labeled by the quantum numbers $j^\pi$ and $(I,S)$ of a static hadron.
The members of the multiplets are labeled by spin and parity quantum numbers $J^P$.
The HQSS multiplet is a singlet with $J=\frac{1}{2}$ if $j=0$ or a doublet with $J=j-\frac{1}{2},j+\frac{1}{2}$ if $j\geq\frac{1}{2}$.
The parity is $P=+\pi$ or $-\pi$ for a hadron with a heavy quark or antiquark.

The $\frac{1}{2}^+$ $\bm{\bar{3}}$-meson triplet corresponds to HQSS doublets of charm mesons with $J^P=0^-,1^-$:
$\bar{D}^{(\ast)}\equiv(\bar{D},\bar{D}^\ast)$ and $\bar{D}_s^{(\ast)}\equiv(\bar{D}_s\bar{D}_s^\ast)$ with $(I,S)=\bigl(\frac{1}{2},0\bigr)$ and $(0,-1)$.
The $0^+$ $\bm{3}$-baryon antitriplet corresponds to charm baryons with $J^P=\frac{1}{2}^+$: $\Lambda_c$ and $\Xi_c$ with $(I,S)=(0,0)$ and $\bigl(\frac{1}{2},-1\bigr)$.
The $1^+$ $\bm{3}$-baryon sextet corresponds to HQSS doublets of charm baryons with $J^P=\frac{1}{2}^+,\frac{3}{2}^+$: $\Sigma_c^{(\ast)}\equiv(\Sigma_c,\Sigma_c^\ast)$, $\Xi_c^{\prime(\ast)}\equiv(\Xi_c^\prime,\Xi_c^{\prime\ast})$, and $\Omega_c^{(\ast)}\equiv(\Omega_c,\Omega_c^\ast)$ with $(I,S)=(1,0)$, $\bigl(\frac{1}{2},-1\bigr)$, and $(0,-2)$.

An $\bm{8}$-hadron with three light quarks is an $\bm{8}$-baryon.
At small $r$, a pentaquark B\nobreakdash-O potential approaches a repulsive color-Coulomb potential offset by the energy of an $\bm{8}$-baryon.
The B-O potentials that are most likely to support bound states are those associated with the ground-state $\bm{8}$-baryons.
The state of an $\bm{8}$-baryon must be completely antisymmetric under exchanges of the three quarks.
The ground-state $\bm{8}$-baryons are expected to be symmetric in the quark positions and therefore have $\pi=+$.
The color state of the three quarks must be a mixed-symmetry octet, so their spin-flavor state must have mixed symmetry.
The ground-state $\bm{8}$-baryons are expected to form four SU(3) multiplets: a $j^\pi=\frac{1}{2}^+$ singlet with $(I,S)=(0,-1)$, a $\frac{1}{2}^+$ octet and a $\frac{3}{2}^+$ octet each consisting of $\bigl(\frac{1}{2},0\bigr)$, $(0,-1)$, $(1,-1)$, and $\bigl(\frac{1}{2},-2\bigr)$, and a $\frac{1}{2}^+$ decuplet consisting of $\bigl(\frac{3}{2},0\bigr)$, $(1,-1)$, $\bigl(\frac{1}{2},-2\bigr)$, and $(0,-3)$.

\textbf{Pentaquark B\nobreakdash-O Potentials.}
The B\nobreakdash-O potentials for hidden-heavy hadrons are labeled by quantum numbers for the symmetries of QCD with static triplet and antitriplet sources. For sources separated by a vector $\bm{r}$, the symmetries are rotations around $\bm{\hat{r}}$, a reflection $R$ through any plane containing $\bm{\hat{r}}$, and $CP$.
The traditional B\nobreakdash-O quantum numbers are $\Lambda_\eta^\epsilon$, where $\Lambda=\lvert\bm{J}\cdot\bm{\hat{r}}\rvert$ with $\bm{J}$ the angular-momentum vector, $\eta=g$ or $u$ for even or odd $CP$, and $\epsilon=+$ or $-$ for even or odd $R$.
For pentaquark B\nobreakdash-O potentials, states with $\eta=g,u$ are equal-energy superpositions with opposite baryon numbers and states with $\epsilon=+,-$ are equal-energy superpositions with opposite $\bm{J}\cdot\bm{\hat{r}}$.
Therefore, the subscript $\eta$ and the superscript $\epsilon$ can both be omitted and the B\nobreakdash-O quantum numbers reduce simply to the half-integer $\Lambda$.
The B\nobreakdash-O potentials are also labeled by flavor quantum numbers $(I,S)$.
The $c\bar{c}$ pentaquarks were discovered through their decays into $J/\psi\, p$, which implies $(I,S)=\bigl(\frac{1}{2},0\bigr)$.
The $c\bar{c}s$ pentaquarks were discovered through their decays into $J/\psi\, \Lambda$, which implies $(I,S)=(0,-1)$.

At large $r$, the B\nobreakdash-O potentials form degenerate multiplets labeled by the quantum numbers $j_\text{B}^{\pi_\text{B}}$, $(I_\text{B},S_\text{B})$ of a $\bm{3}$-baryon and $j_\text{M}^{\pi_\text{M}}$, $(I_\text{M},S_\text{M})$ of a $\bm{\bar{3}}$-meson, which can be alternatively specified by the HQSS multiplets of heavy baryons and heavy mesons.
Each multiplet of B\nobreakdash-O potentials can be decomposed into sub-multiplets labeled by $j^\pi$ with $j=\lvert j_\text{B}-j_\text{M}\rvert,\dots,j_\text{B}+j_\text{M}$ and $\pi=\pi_\text{B}\pi_\text{M}$.
We label these potentials by $(j^\pi)\Lambda$ with $\Lambda=\frac{1}{2},\dots,j$.
Their isospin is $I=\lvert I_\text{B} - I_\text{M}\rvert,\dots,I_\text{B}+I_\text{M}$ and their strangeness is $S=S_\text{B}+S_\text{M}$.
The B\nobreakdash-O potentials with $(I,S)=\bigl(\frac{1}{2},0\bigr)$ and $(0,-1)$ that have the lowest energy at large $r$ are listed in Table~\ref{tab:3bar3pot}.

\begin{table}
\caption{\label{tab:3bar3pot}The pairs of a ground-state $\bm{3}$-baryon and a ground-state $\bm{\bar{3}}$-meson with $(I,S)=\bigl(\frac{1}{2},0\bigr)$ and $(0,-1)$ and the associated B-O potentials $(j^\pi)\Lambda$ at large $r$.
}
\begin{ruledtabular}
\begin{tabular}{cccl}
Static-hadron pair & $I$ & $S$	& $(j^\pi)\Lambda$	\\
\hline
$\Lambda_c \bar{D}^{(\ast)}$	& $\frac{1}{2}$	& 0	& $\bigl(\frac{1}{2}^+\bigr)\frac{1}{2}$ \\
$\Sigma_c^{(\ast)} \bar{D}^{(\ast)}$	& $\frac{1}{2}$	& 0	& $\bigl(\frac{3}{2}^+\bigr)\frac{1}{2}$, $\bigl(\frac{3}{2}^+\bigr)\frac{3}{2}$, $\bigl(\frac{1}{2}^{+\prime}\bigr)\frac{1}{2}$  \\
$\Lambda_c \bar{D}_s^{(\ast)}$	& 0	& $-1\hphantom{-}$	& $\bigl(\frac{1}{2}^+\bigr)\frac{1}{2}$ \\
$\Xi_c \bar{D}^{(\ast)}$	& 0	& $-1\hphantom{-}$	& $\bigl(\frac{1}{2}^{+\prime}\bigr)\frac{1}{2}$  \\
$\Xi_c^{\prime(\ast)}\bar{D}^{(\ast)}$	& 0	& $-1\hphantom{-}$	& $\bigl(\frac{3}{2}^+\bigr)\frac{1}{2}$, $\bigl(\frac{3}{2}^+\bigr)\frac{3}{2}$, $\bigl(\frac{1}{2}^{+\prime\prime}\bigr)\frac{1}{2}$  \\
\end{tabular}
\end{ruledtabular}
\end{table}

At small $r$, the B\nobreakdash-O potentials form degenerate multiplets labeled by the quantum numbers $j^\pi$ and $(I,S)$ of an $\bm{8}$-baryon.
We label these potentials by $(j^\pi)\Lambda$ with $\Lambda=\frac{1}{2},\dots,j$.
Their flavor is that of the $\bm{8}$-baryon.
The B\nobreakdash-O potentials with $(I,S)=\bigl(\frac{1}{2},0\bigr)$ and $(0,-1)$ that have the lowest energy at small $r$ are listed in Table~\ref{tab:8pot}.

\begin{table}
\caption{\label{tab:8pot} The ground-state $\bm{8}$-baryons with $(I,S)=\bigl(\frac{1}{2},0\bigr)$ and $(0,-1)$ and the associated B-O potentials $(j^\pi)\Lambda$ at small $r$.
}
\begin{ruledtabular}
\begin{tabular}{cccl}
$\bm{8}$-baryon	& $I$ & $S$	& $(j^\pi)\Lambda$	\\
\hline
$\frac{1}{2}^+$	& $\frac{1}{2}$	& 0	& $\bigl(\frac{1}{2}^+\bigr)\frac{1}{2}$ \\
$\frac{3}{2}^+$	& $\frac{1}{2}$	& 0	& $\bigl(\frac{3}{2}^+\bigr)\frac{1}{2}$, $\bigl(\frac{3}{2}^+\bigr)\frac{3}{2}$ \\
$\frac{1}{2}^+$	& 0	& $-1\hphantom{-}$	& $\bigl(\frac{1}{2}^+\bigr)\frac{1}{2}$ \\
$\hphantom{^\prime}\frac{1}{2}^{+\prime}$	& 0	& $-1\hphantom{-}$	& $\bigl(\frac{1}{2}^{+\prime}\bigr)\frac{1}{2}$ \\
$\frac{3}{2}^+$	& 0	& $-1\hphantom{-}$	& $\bigl(\frac{3}{2}^+\bigr)\frac{1}{2}$, $\bigl(\frac{3}{2}^+\bigr)\frac{3}{2}$ \\
\end{tabular}
\end{ruledtabular}
\end{table}

Since the spectrum of QCD with two static color sources must be a smooth function of $r$, the B\nobreakdash-O potentials associated with a $\bm{3}$-baryon and $\bm{\bar{3}}$-meson at large $r$ must connect smoothly to the B\nobreakdash-O potentials associated with an $\bm{8}$-baryon at small $r$.
The $(j^\pi)\Lambda$ potential with the nth lowest energy at large $r$ must connect to the B\nobreakdash-O potential with the same $\Lambda$ and $(I,S)$ and the nth lowest energy at small $r$.
The simplest possibility is that the quantum numbers $j^\pi$ at small $r$ coincide with those at large $r$.
Under this assumption, we proceed to connect the B\nobreakdash-O potentials in Table~\ref{tab:3bar3pot} to those in Table~\ref{tab:8pot}.
We then use these connections together with experimental evidence to infer the potentials that may support bound states.

First, we examine B\nobreakdash-O potentials with $(I,S)=\bigl(\frac{1}{2},0\bigr)$. 
The $\Lambda_c \bar{D}^{(\ast)}$ potential $\bigl(\frac{1}{2}^+\bigr)\frac{1}{2}$ must connect to the potential for the $\frac{1}{2}^+$ $\bm{8}$-baryon.
The two $\Sigma_c^{(\ast)} \bar{D}^{(\ast)}$ potentials $\bigl(\frac{3}{2}^+\bigr)\Lambda$ must connect to the two potentials for the $\frac{3}{2}^+$ $\bm{8}$-baryon.
The $\Sigma_c^{(\ast)} \bar{D}^{(\ast)}$ potential $\bigl(\frac{1}{2}^{+\prime}\bigr)\frac{1}{2}$ must connect to the potential for an orbitally excited $\bm{8}$-baryon, since Table~\ref{tab:8pot} does not have a second $\bm{8}$-baryon with $j^\pi=\frac{1}{2}^+$ and $(I,S) = \bigl(\frac{1}{2},0\bigr)$.
The lack of evidence for $c \bar{c}$ pentaquarks near the $\Lambda_c \bar{D}$ and $\Lambda_c \bar{D}^\ast$ thresholds suggests that the $\bigl(\frac{1}{2}^+\bigr)\frac{1}{2}$ potential does not support bound states.
One of the three observed $c \bar{c}$ pentaquarks is near the $\Sigma_c \bar{D}$ threshold and the other two are near the $\Sigma_c \bar{D}^\ast$ threshold.
This suggests that either the $\bigl(\frac{3}{2}^+\bigr)\Lambda$ potentials or the $\bigl(\frac{1}{2}^{+\prime}\bigr)\frac{1}{2}$ potential or both support bound states.
It is implausible that the $\bigl(\frac{1}{2}^{+\prime}\bigr)\frac{1}{2}$ potential supports bound states, since it connects to the potential of an orbitally-excited $\bm{8}$-baryon.
In this case, only the $\bigl(\frac{3}{2}^+\bigr)\Lambda$ potentials would support bound states.

Next, we examine B\nobreakdash-O potentials with $(I,S)=(0,-1)$.
The $\Lambda_c \bar{D}_s^{(\ast)}$ potential $\bigl(\frac{1}{2}^+\bigr)\frac{1}{2}$ must connect to the potential for the $\frac{1}{2}^+$ $\bm{8}$-baryon.
The $\Xi_c \bar{D}^{(\ast)}$ potential $\bigl(\frac{1}{2}^{+\prime}\bigr)\frac{1}{2}$ must connect to the potential for the $\frac{1}{2}^{+\prime}$ $\bm{8}$-baryon.
The $\frac{1}{2}^+$ and  $\frac{1}{2}^{+\prime}$ $\bm{8}$-baryons are orthogonal linear combinations of the ground-state $\bm{8}$-baryons in the SU(3) singlet and the SU(3) octet.
The two $\Xi_c^{\prime(\ast)} \bar{D}^{(\ast)}$ potentials $\bigl(\frac{3}{2}^+\bigr)\Lambda$ must connect to the two potentials for the $\frac{3}{2}^+$ $\bm{8}$-baryon.
The $\Xi_c^{\prime(\ast)} \bar{D}^{(\ast)}$ potential $\bigl(\frac{1}{2}^{+\prime\prime}\bigr)\frac{1}{2}$ must connect to the potential for an orbitally excited $\bm{8}$-baryon, since Table~\ref{tab:8pot} does not have a third $\bm{8}$-baryon with $j^\pi=\frac{1}{2}^+$ and $(I,S) = (0,-1)$.
The lack of evidence for $c \bar{c}s$ pentaquarks near the $\Lambda_c \bar{D}_s$ and $\Lambda_c \bar{D}_s^\ast$ thresholds suggests that the $\bigl(\frac{1}{2}^+\bigr)\frac{1}{2}$ potential does not support bound states.  
The two observed $c \bar{c}s$ pentaquarks are near the $\Xi_c \bar{D}$ and $\Xi_c \bar{D}^\ast$ thresholds.
This suggests that the $\bigl(\frac{1}{2}^{+\prime}\bigr)\frac{1}{2}$ potential supports bound states.

\textbf{Pentaquark Energies.}
At leading order in $1/m_Q$, hidden-heavy pentaquarks in B\nobreakdash-O potentials $(j^\pi)\Lambda$ form degenerate HQSS multiplets whose members are labeled by spin and parity quantum numbers $J^P$.
The ground-state multiplet in B\nobreakdash-O potentials $(j^\pi)\Lambda$ consists of a heavy-quark spin-singlet ($S_{Q\bar{Q}}=0$) state with $J=j$ and two or three heavy-quark spin-triplet ($S_{Q\bar{Q}}=1$) states with $J$ ranging from $\lvert j - 1\rvert$ to $j + 1$. The parity is $P=-\pi$.

The first-order correction in $1/m_Q$ to the B\nobreakdash-O potentials depends on $\bm{r}$ and the spin vectors $\bm{S}_Q$ and $\bm{S}_{\bar{Q}}$ of the heavy quark and antiquark.
It includes short-distance terms that give spin splittings to the $\bm{8}$-baryons and long-distance terms that give spin splittings to well-separated heavy hadrons.
At large distances, the correction is a spin-splitting potential $V_\text{SS}$ that does not depend on $\bm{r}$.
For the pentaquark B\nobreakdash-O potentials in Table~\ref{tab:3bar3pot},
\begin{equation}
V_\text{SS}(\bm{S}_Q, \bm{S}_{\bar{Q}}) = \tfrac{2}{3} \Delta_\text{B} \bm{j}_\text{B} \cdot \bm{S}_Q + \Delta_\text{M} \bm{j}_\text{M} \cdot \bm{S}_{\bar{Q}},
\end{equation}
where $\bm{j}_\text{B}$ and $\bm{j}_\text{M}$ are the spins of the $\bm{3}$-baryon and $\bm{\bar{3}}$-meson and $\Delta_\text{B}$ and $\Delta_\text{M}$ are the spin splittings of the heavy baryons and heavy mesons.
A simple approximation to the spin splittings of the pentaquarks can be obtained using first-order perturbation theory in $V_\text{SS}$.
The spin splittings are calculated by diagonalizing the matrix $V_\text{SS}$ for the states in the HQSS multiplet.
We set the orbital angular momentum of the $Q\bar{Q}$ pair to zero since the ground state in the B\nobreakdash-O potentials is expected to be mostly $S$-wave.

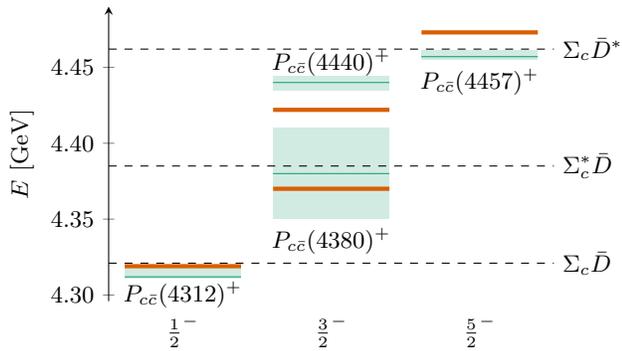
\begin{figure}
\tikzsetnextfilename{c-c-bar-levels}
\begin{tikzpicture}
	\pgfplotsset{
	/pgfplots/error bars/draw error bar/.code 2 args={%
		\node[inner sep=0pt] (a) at #1 {};
		\node[inner sep=0pt] (b) at #2 {};
		\path[draw=none, fill, opacity=.2] ($(a) - (22pt,0)$) rectangle ($(b) + (22pt,0)$);
		}
	}
	\pgfplotstableread{
		x	y
		0	4.319
		1	4.370
		1	4.422
		2	4.473
	}{\nonstrangetheor}
	\pgfplotstableread{
		x 	y 	ey+ 	ey-
		0	4.312	0.007	0.001
		1	4.440	0.004	0.005
		2	4.457	0.004	0.002
	}{\nonstrangeexp}
	\begin{axis}[
		width=7.5cm,
		height=5.5cm,
		only marks,
		black,
		clip mode=individual,
		ylabel=$E$,
		y unit=eV,
		y SI prefix=giga,
		axis y line=left,
		axis x line=bottom,
		ymin=4.296,
		ymax=4.49,
		xmin=-.5,
		xmax=2.5,
		x axis line style={draw=none},
		x tick style={draw=none},
		xtick={0, 1, 2},
		xticklabels={$\frac{1}{2}^{-}$, $\frac{3}{2}^{-}$, $\frac{5}{2}^{-}$},
		ytick distance=0.05,
		yticklabel style={/pgf/number format/.cd,fixed,zerofill,precision=2},
		cycle list/Dark2
		]
		\addplot+[mark=-,mark size=22pt,
			error bars/.cd, y dir=both, y explicit]
			table[y error plus=ey+, y error minus=ey-] {\nonstrangeexp};
		\addplot+[mark=-,ultra thick,mark size=22pt] 
			table {\nonstrangetheor};
		\begin{scope}[dashed]
			\draw (-.5, 4.321) -- (2.5, 4.321)
			node[anchor=west] {$\Sigma_c\bar{D}$};
			\draw (-.5, 4.385) -- (2.5, 4.385)
			node[anchor=west] {$\Sigma_c^\ast\bar{D}$};
			\draw (-.5, 4.462) -- (2.5, 4.462)
			node[anchor=west] {$\Sigma_c\bar{D}^\ast$};
		\end{scope}
		\node[anchor=north,inner sep=1pt] at (0, 4.311) {$P_{c\bar{c}}(4312)^+$};
		\node[anchor=south, inner sep=0pt] at (1, 4.444) {$P_{c\bar{c}}(4440)^+$};
		\node[anchor=north] at (2, 4.455) {$P_{c\bar{c}}(4457)^+$};
	\end{axis}
\end{tikzpicture}
\caption{\label{fig:ccbardiagram}
Energy levels for the ground-state quartet of $c \bar{c}$ pentaquarks in the $\bigl(\frac{3}{2}^+\bigr)\Lambda$ potentials with $(I,S)=\bigl(\frac{1}{2},0\bigr)$ (thick bars) versus the measured masses (shaded boxes with thin lines).}
\end{figure}

We identify the three observed $c \bar{c}$ pentaquarks as members of the ground-state quartet in the $\bigl(\frac{3}{2}^+\bigr) \Lambda$ potentials that approach the $\Sigma_c^{(\ast)} \bar{D}^{(\ast)}$ threshold at large $r$.
We predict the $J^P$ quantum numbers for $P_{c\bar{c}}(4312)^+$, $P_{c\bar{c}}(4440)^+$, and $P_{c\bar{c}}(4457)^+$ to be $\frac{1}{2}^-$, $\frac{3}{2}^-$, and $\frac{5}{2}^-$, respectively.
We also predict an additional $\frac{3}{2}^-$ state.
We identify the energy of the multiplet in the absence of spin splittings with the spin-weighted average of the central values of the measured masses, which we calculate using their predicted quantum numbers.
We approximate the spin splittings by first-order perturbation theory in $V_\text{SS}$. We determine $\Delta_\text{B}=66.5$ MeV from the mass splitting between $\Sigma_c^\ast$ and $\Sigma_c$ and $\Delta_\text{M}=142.1$~MeV from the mass splitting between $\bar{D}^\ast$ and $\bar{D}$.
The energy level diagram is illustrated in Figure~\ref{fig:ccbardiagram}.
The additional $\frac{3}{2}^-$ state is predicted to be near the $\Sigma_c^\ast \bar{D}$ threshold.
We denote it by $P_{c\bar{c}}(E)^+$ with $E$ near 4385.

We identify the two observed $c \bar{c} s$ pentaquarks as members of the ground-state triplet for the $\bigl(\frac{1}{2}^{+\prime}\bigr)\frac{1}{2}$ potential that approaches the $\Xi_c \bar{D}^{(\ast)}$ threshold at large $r$.
The third pentaquark is predicted to have a mass close to that of $P_{c\bar{c}s}(4459)^0$.
The quantum numbers of $P_{c\bar{c}s}(4338)^0$ are $\frac{1}{2}^-$, in agreement with experiment  \cite{LHCb22}.
We predict the quantum numbers of $P_{c\bar{c}s}(4459)^0$ and the third pentaquark to be $\frac{1}{2}^-$ and $\frac{3}{2}^-$ or $\frac{3}{2}^-$ and $\frac{1}{2}^-$.
We identify the energy of the multiplet in the absence of spin splittings with the spin-weighted average of the central values of the measured masses, which we calculate using the predicted quantum numbers for $P_{c\bar{c}s}(4459)^0$ and the third pentaquark with the same mass.
We approximate the spin splittings by first-order perturbation theory in $V_\text{SS}$. 
We determine $\Delta_\text{M}=142.1$~MeV from the mass splitting between $\bar{D}^\ast$ and $\bar{D}$ and we set $\Delta_\text{B}=0$ since $\Xi_c$ is a HQSS singlet.
The energy level diagram is illustrated in Figure~\ref{fig:ccbarsdiagram}.

\begin{figure}
\tikzsetnextfilename{c-c-bar-s-levels}
\begin{tikzpicture}
	\pgfplotsset{
		/pgfplots/error bars/draw error bar/.code 2 args={%
			\node[inner sep=0pt] (a) at #1 {};
			\node[inner sep=0pt] (b) at #2 {};
			\path[draw=none, fill, opacity=.2] ($(a) - (22pt,0)$) rectangle ($(b) + (22pt,0)$);
		}
	}
	\pgfplotstableread{
		x	y
		0	4.322
		0	4.464
		1	4.464
	}{\strangetheor}
	\pgfplotstableread{
		x 	y 	ey+ 	ey-
		0	4.338	0.001	0.001
		0	4.459	0.006	0.003
		1	4.459	0.006	0.003
	}{\strangeexp}
	\begin{axis}[
		width=5.7cm,
		height=5.5cm,
		only marks,
		black,
		clip mode=individual,
		ylabel=$E$,
		y unit=eV,
		y SI prefix=giga,
		axis y line=left,
		axis x line=bottom,
		ymin=4.296,
		ymax=4.49,
		xmin=-.5,
		xmax=1.5,
		x axis line style={draw=none},
		x tick style={draw=none},
		xtick={0, 1, 2},
		xticklabels={$\frac{1}{2}^{-}$, $\frac{3}{2}^{-}$},
		ytick distance=0.05,
		yticklabel style={/pgf/number format/.cd,fixed,zerofill,precision=2},
		cycle list/Dark2
		]
		\addplot+[mark=-,mark size=22pt,
			error bars/.cd, y dir=both, y explicit]
			table[y error plus=ey+, y error minus=ey-] {\strangeexp};
		\addplot+[mark=-,ultra thick,mark size=22pt] 
			table {\strangetheor};
		\begin{scope}[dashed]
			\draw (-.5, 4.336) -- (1.5, 4.336)
			node[anchor=west] {$\Xi_c\bar{D}$};
			\draw (-.5, 4.478) -- (1.5, 4.478)
			node[anchor=west] {$\Xi_c\bar{D}^\ast$};
		\end{scope}
		\node[anchor=south] at (0, 4.339) {$P_{c\bar{c}s}(4338)^0$};
		\node[anchor=north] at (0, 4.456) {$P_{c\bar{c}s}(4459)^0?$};
		\node[anchor=north] at (1, 4.456) {$P_{c\bar{c}s}(4459)^0?$};
	\end{axis}
\end{tikzpicture}
\caption{\label{fig:ccbarsdiagram}
Energy levels for the ground-state triplet of $c \bar{c}s$ pentaquarks in the $\bigl(\frac{1}{2}^{+\prime}\bigr)\frac{1}{2}$ potential with $(I,S)=(0,-1)$ (thick bars) versus the measured masses (shaded boxes with thin lines).
The predicted quantum numbers of $P_{c\bar{c}s}(4459)^0$ are either $\frac{1}{2}^-$ or $\frac{3}{2}^-$.}
\end{figure}

\textbf{Pentaquark Decays.}
The decays of hidden-heavy pentaquarks into heavy-hadron pairs could produce large partial widths unless they are suppressed.
The observed pentaquarks have surprisingly narrow decay widths.
It is therefore important to identify a suppression mechanism for the kinematically allowed decays into charm-hadron pairs for each narrow pentaquark.
The decay into a constituent heavy-hadron pair can be calculated using first-order perturbation theory in $V_\text{SS}$.
The decay rate is the product of the probability for that heavy-hadron pair as a constituent, the square of the matrix element of $V_\text{SS}$, and a phase space factor.
The decay into a lower-energy heavy-hadron pair can proceed through a transition between B\nobreakdash-O potentials.
Model-independent relations for these decays can be obtained using the techniques introduced in Ref.~\cite{Braa24a}.

The $c\bar{c}$ pentaquarks can decay into $\Lambda_c \bar{D}$ and $\Lambda_c \bar{D}^\ast$ through a transition between the $\bigl(\frac{3}{2}^+\bigr) \frac{1}{2}$ and $\bigl(\frac{1}{2}^+\bigr) \frac{1}{2}$ potentials in Table~\ref{tab:3bar3pot}.
The relations of Ref.~\cite{Braa24a} indicate that these decays have $D$-wave suppression, which is a remarkable prediction of the B\nobreakdash-O framework.
$P_{c\bar{c}}(4440)^+$ and $P_{c\bar{c}}(4457)^+$ are kinematically allowed to decay into the constituents $\Sigma_c\bar{D}$ and $\Sigma_c^\ast\bar{D}$.
The fourth $c\bar{c}$ pentaquark $P_{c\bar{c}}(E)^+$ is kinematically allowed to decay into $\Sigma_c\bar{D}$ and also into $\Sigma_c^\ast\bar{D}$ if $E>4385$.
The only decays that can proceed through an $S$-wave channel are $P_{c\bar{c}}(E)^+$ and  $P_{c\bar{c}}(4440)^+$ into $\Sigma_c^\ast\bar{D}$.
The probabilities for $\Sigma_c^\ast\bar{D}$ in $P_{c\bar{c}}(E)^+$ and  $P_{c\bar{c}}(4440)^+$, which are determined by $\Delta_\text{B}/\Delta_\text{M}$, are 0.611 and 0.056.
The tiny $\Sigma_c^\ast\bar{D}$ probability for $P_{c\bar{c}}(4440)^+$ is a remarkable prediction of the B\nobreakdash-O framework.
The matrix elements of $V_\text{SS}$ factor out of the ratio.
If $E> 4385$, the ratio of the decay rates is the ratio of  the phase-space factors multiplied by the probability ratio 10.9.

The $c\bar{c}s$ pentaquarks can decay into $\Lambda_c\bar{D}_s$ and $\Lambda_c\bar{D}^\ast_s$ through a transition between the $\bigl(\frac{1}{2}^{+\prime}\bigr) \frac{1}{2}$ and $\bigl(\frac{1}{2}^+\bigr) \frac{1}{2}$ potentials in Table~\ref{tab:3bar3pot}.
Generalizing the model-independent relations of Ref.~\cite{Braa24a} to include superpositions of $S_{Q\bar{Q}}=0$ and 1, we predict
\begin{equation}
\frac{\Gamma[P_{c\bar{c}s}(4459)^0  \to  \Lambda_c^+\, D_s^{\ast  -}]}{\Gamma[P_{c\bar{c}s}(4338)^0  \to \Lambda_c^+\, D_s^-]} = 0.85^{+0.04}_{-0.02} \,.
\label{eq:gamma4459/4338}
\end{equation}
This prediction is just the phase-space ratio and its uncertainty comes from the pentaquark masses.
$P_{c\bar{c}s}(4459)^0$ is also kinematically allowed to decay into the constituents $\Xi_c\bar{D}$, but the probability for the $\Xi_c\bar{D}$ component is zero indicating $D$-wave suppression.
The ratio of the measured widths of $P_{c\bar{c}s}(4459)^0$ and $P_{c\bar{c}s}(4338)^0$ is $2.5^{+1.6}_{-1.4}$, which is compatible with the ratio in Equation~\eqref{eq:gamma4459/4338}.
The transition amplitude between the $\bigl(\frac{1}{2}^{+\prime}\bigr) \frac{1}{2}$ and $\bigl(\frac{1}{2}^+\bigr) \frac{1}{2}$ potentials is proportional to their flavor overlap.
The flavors of $\Xi_c\bar{D}^{(\ast)}$ and $\Lambda_c\bar{D}_s^{(\ast)}$ are orthogonal, so the transition amplitude approaches zero at large $r$.
The transition amplitude at small $r$ is determined by the mixing between the SU(3) singlet and the $(0,-1)$ member of the SU(3) octet of $\bm{8}$-baryons with $j^\pi=\frac{1}{2}^+$.
The small width for the $c \bar{c} s$ pentaquarks into $\Lambda_c\bar{D}_s$ and $\Lambda_c\bar{D}_s^\ast$ could be explained by the decays being dominated by a region of $r$ where the flavor overlap is small.

\textbf{Molecular Picture.}
The molecular picture corresponds in the B\nobreakdash-O framework to the assumption that all the B\nobreakdash-O potentials that approach the same static-hadron-pair threshold at large $r$ support bound states.
If these potentials are the same as those associated with a single low-energy $\bm{8}$-baryon at small $r$, the B\nobreakdash-O approximation will be qualitatively similar to the molecular picture.
An example is $c \bar{c} s$ pentaquarks with $(I,S) = (0,-1)$.
If there are additional  B\nobreakdash-O potentials that approach the static-hadron-pair threshold at large $r$, the B\nobreakdash-O approximation may be qualitatively different from the molecular picture.
An example is $c \bar{c}$ pentaquarks with $(I,S)=\bigl(\frac{1}{2},0\bigr)$.
The molecular picture predicts seven ground-state $c \bar{c}$ pentaquarks compared to the B\nobreakdash-O prediction of four.
A recent study has considered an alternative B\nobreakdash-O approximation that reproduces the molecular picture by ignoring the requirement that the $\frac{1}{2}^{+\prime}$ potential must connect to that for an orbitally excited $\bm{8}$-baryon \cite{Bra25}.

If the $\frac{1}{2}^+$ $\bm{8}$-baryon with $(I,S) = \bigl( \frac{3}{2},0 \bigr)$ has low enough energy, there could be isospin-$\frac{3}{2}$ $c \bar{c}$ pentaquarks.
They could be discovered through decays into $J/\psi\, \Delta^{++}$, with $\Delta^{++}$ decaying into $p\, \pi^+$ which results in two charged hadrons.
The  $(I,S) = \bigl( \frac{3}{2},0 \bigr)$ potentials that approach the $\Sigma_c^{(\ast)} \bar{D}^{(\ast)}$  threshold at large $r$ are $\bigl( \frac{1}{2}^+ \bigr)\frac{1}{2}$ and $\bigl( \frac{3}{2}^+ \bigr)\Lambda$.
The only low-energy $\bm{8}$-baryon with $(I,S) = \bigl( \frac{3}{2},0 \bigr)$ is $\frac{1}{2}^+$ and the associated potential at small $r$ is $\bigl( \frac{1}{2}^+ \bigr)\frac{1}{2}$.
The B\nobreakdash-O approximation predicts that $\bigl( \frac{1}{2}^+ \bigr)\frac{1}{2}$ is the only potential that may support bound states.
The ground-state multiplet would consist of three $c\bar{c}$ pentaquarks compared to seven predicted by the molecular picture.

\textbf{Summary.}
We have applied the pattern of exotic hidden-heavy hadrons proposed in Ref.~\cite{Braa24b} to hidden-heavy pentaquarks.
We have used the B\nobreakdash-O symmetries to connect hadron-pair potentials at long distances with $\bm{8}$-hadron potentials at short distances.
We have used these connections together with experimental evidence to infer which B\nobreakdash-O potentials may support bound states.
The three observed $c \bar{c}$ pentaquarks belong to the HQSS quartet for the $c \bar{c}$ ground state in $\bigl(\frac{3}{2}^+\bigr)\Lambda$ potentials.
The two observed $c \bar{c} s$ pentaquarks belong to the HQSS triplet for the $c \bar{c}$ ground state in a $\bigl(\frac{1}{2}^{+\prime}\bigr)\frac{1}{2}$ potential.
Our most surprising prediction is that $P_{c\bar{c}}(4457)^+$ has quantum numbers $\frac{5}{2}^-$, which agrees with only a few previous predictions \cite{Ani15,Shi21,Wan22}.
Most previous studies have predicted $\frac{1}{2}^-$ or $\frac{3}{2}^-$; see, for instance, Refs.~\cite{Du19,Pav19,Xia19,Liu20,Gir21,Bur22,Den22,Li23,Xu23,Cly24,Pen24} and references therein.
Our prediction could be verified by resolving the angular distribution of the decay products in the discovery channel $P_{c\bar{c}}(4457)^+ \to J/\psi\,p$.
If the quantum numbers of $P_{c\bar{c}}(4457)^+$ are $\frac{1}{2}^-$ or $\frac{3}{2}^-$, the decay can proceed in $S$-wave.
Our assignment $\frac{5}{2}^-$ rules out $S$-wave, and we instead predict that the decay $P_{c\bar{c}}(4457)^+ \to J/\psi\,p$ will be dominated by $D$-wave.

According to the proposal in Ref.~\cite{Braa24b}, the pattern of  exotic hidden-heavy hadrons is largely determined by the spectrum of $\bm{8}$-hadrons in QCD.
That spectrum has been predicted using various constituent models \cite{ATLAS19review}.
The discoveries of the exotic hidden-heavy hadrons provide a powerful incentive for definitive calculations of the spectrum of $\bm{8}$-hadrons using lattice QCD or lattice NRQCD.

Definitive predictions of the properties of hidden-heavy hadrons will require lattice QCD calculations of the relevant B\nobreakdash-O potentials.
Until those potentials are available, the best one can do is develop simple models for the potentials consistent with the constraints from BOEFT.
Solutions of the Schr\"odinger equations with those model potentials and with spin splittings of heavy hadrons treated nonperturbatively should give more accurate predictions of the spectrum of the exotic hidden-heavy hadrons.
The experimental discoveries of additional hidden-heavy tetraquarks and pentaquarks consistent with those predictions would confirm that the pattern of the exotic hidden-heavy hadrons in QCD is finally understood.

\bigskip

\begin{acknowledgments}
We thank T. Skwarnicki for valuable comments.
This work was supported in part by the U.S. Department of Energy under grant DE-SC0011726. This work contributes to the goals of the US DOE ExoHad Topical Collaboration, Contract DE-SC0023598.
\end{acknowledgments}

\bibliography{bibliography}
	
\end{document}